\documentstyle[aps,prl,multicol,epsfig,amsfonts]{revtex}

\begin{document}

\title{Holonomic quantum computation with neutral atoms}

\author{A. Recati$^{1,2,3}$, T. Calarco$^{1,2}$, P. Zanardi$^4$, J.I. Cirac$^5$,
and P. Zoller$^2$}
\address{$^1$ ECT$^*$, I-38050 Villazzano (TN) and Istituto Nazionale di
Fisica della Materia, I-38050 Povo, Italy\\
$^2$ Institut f\"ur Theoretische Physik, Universit\"at Innsbruck,
A-6020 Innsbruck, Austria \\
$^3$ Dipartimento di Fisica, Universit\`a di Trento, I-38050 Povo, Italy \\
$^4$ Institute for Scientific Interchange (ISI) Foundation,
I-10133 Torino, Italy \\
$^5$ Max-Planck Institut f\"ur Quantenoptik, Hans-Kopfermann Str. 1, D-85748 Garching, Germany}

\maketitle

\begin{abstract}
We propose an all-geometric implementation of quantum computation
using neutral atoms in cavity QED. We show how to perform generic
single- and two-qubit gates, the latter by encoding a two-atom
state onto a single, many-level atom. We compare different strategies to
overcome limitations due to cavity imperfections.
\end{abstract}

\begin{multicols}{2}
The  standard paradigm of quantum computation (QC) \cite{QC} is a
dynamical one: in order to manipulate the  quantum state of
systems encoding information, local interactions between low-dimensional 
subsystems (qubits) are switched on and off in such a
way  to enact a sequence of quantum gates. On the other hand,
ever since the discovery of the Berry's phase, it has been
recognized that quantum evolutions, besides dynamical
contributions, can display purely geometrical effects \cite{SHWI}.
The latter, in view of their  very geometric-topological nature,
show an inherent stability against some local perturbations.
It is therefore a natural and intriguing question to ask whether
one could take advantage of this geometric features to the aim of
processing quantum information. Indeed one would expect the
above mentioned robustness to  result in a resilience against
some kinds of errors. In other terms a geometry-based strategy for
quantum manipulations is expected to have some built-in
fault-tolerant features \cite{presk}. In the context of NMR
\cite{a} and Josephson junction based quantum computing \cite{b}
it has been show how to use geometrical phases to implement a
two-qubit gate that, along with the dynamically generated
one-qubit gates, is universal.\\
To achieve an all-geometrical implementation of quantum
computation one is  led  to consider  more sophisticated, i.e.,
non-Abelian, structures. This has been originally done in Ref.
\cite{HQC}, where the so called holonomic approach to quantum
computation (HQC) has been introduced. Quite recently a proposal
for implementing an HQC scheme with trapped ions, feasible with
the current technology, has been put forward \cite{IgnaLu}.
In this  paper we discuss an implementation proposal  for HQC by
means  of neutral atoms in cavity QED. This is to some extent related, 
at least regarding single-qubit operations, to the proposal \cite{IgnaLu}. 
We shall show how to perform generic single-qubit gates by using a single atom. For
realizing universal two-qubit gate  a mapping of a two-qubit state
onto a single many-level atom will be used. Finally we shall
propose  a strategy to overcome limitations due to cavity
imperfections.

\section{Holonomic Quantum Computation}

We now briefly recall the basics of HQC \cite{PaoloPachos}. In
the HQC paradigm  information is encoded in an $n$-fold degenerate
eigenspace $\cal C$ of a Hamiltonian $H(\lambda_0)$  belonging
to a $d$-dimensional parametric family of isodegenerate
Hamiltonians $\{H(\lambda)\}. $ The $\lambda$'s represent
parameters that  are supposed to be  controllable  in the given
experimental situation. The manipulations of the codewords in
$\cal C$  are enacted by driving the control parameters along
loops $\chi$  in an adiabatic fashion. In this way an initial
preparation $|\Psi_0\rangle\in{\cal C}$ evolves, up to an overall
dynamical phase, according to the rule $|\Psi_0\rangle\in{\cal
C}\mapsto U(\chi)_A\,|\Psi_0\rangle$ where $U(\chi)_A={\bf
P}\exp \int_\chi A$ is the { holonomy} associated with $\chi$
by the the $u(n)$-valued connection  $A=\sum_{\mu=1}^d A_\mu
\,d\lambda_\mu$. One finds \cite{Wilczek}
\begin{equation}
A_\mu^{\alpha\beta}=\langle\psi^\alpha(\lambda)|\frac{\partial}{\partial\lambda_\mu}|\psi^\beta(\lambda)\rangle,
\end{equation}
where $\{|\psi^\alpha(\lambda)\rangle\}_{\alpha=1}^n$ denotes an
orthonormal basis of the degenerate eigenspace ${\cal C}$. The set
of all possible holonomies, obtained by taking all possible
$\chi$'s, is a subgroup,  known as the holonomy group,  of  the
group $U(n)$ of unitary transformations over $\cal C$.

When the holonomy group  coincides with the whole $U(n)$ one can
perform { universal} QC over $\cal C$ by resorting to geometrical
means only. This irreducibility condition can be easily  stated
in terms of the { curvature} $2$-form $F$ associated with  $A$ by
the relation $F_{\mu\nu}=\partial_\mu A_\nu -\partial_\nu A_\mu
-[A_\mu,\,A_\nu]$ \cite{Nakahara}. The number of linearly
independent $F_{\mu\nu}$'s gives a lower bound to the dimension of  the holonomy
group \cite{notice}. The curvature form $F$ encodes for the
non-trivial geometric features of the global bundle of
$n$-dimensional quantum codes over the  manifold of control
parameters. Flat bundles, i.e., with $F=0$, have no computational
power.

\section{Dark states in (N+1)-level system}

\begin{figure}
\centerline{\epsfig{file=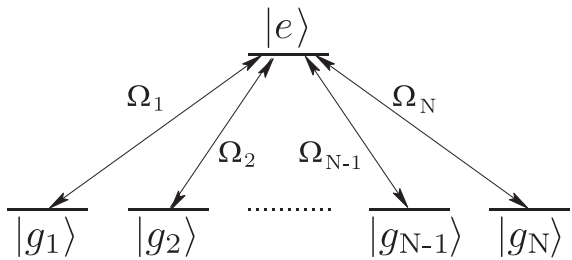,width=7.5cm}}
\caption{Internal level scheme of the atoms inside the cavity,
with the coupling needed to realize the Hamiltonian
Eq. (\ref{eq:HN}).} \label{fig:levels}
\end{figure}

Following the previous recipe we will give the expression of a parametric
Hamiltonian which turns out to be suitable to achieve holonomic quantum 
computation. After that we will show how such Hamiltonian can be implemented using neutral
atoms in an optical resonator. We can, in this way, give a precise physical meaning 
to the abstract objects of the HQC paradigm. The idea relies on the concept of the 
adiabatic passage via dark states. In the simplest case (a.k.a. $\Lambda$-system) we have 
two states (ground states) which are not directly connected, i.e., in the Hamiltonian 
 describing the
system there are no terms which couple these two states -- considering 
an atom interacting with the electromagnetic field there is no single-photon
transition between the states we are interested in. However they are independently 
coupled to a third state (excited state) in a tunable way.
By changing in an adiabatic way the coupling constants, it is possible to create
coherent superpositions of the two ground states and to pass from one to the other 
without populating the excited state \cite{RMPColl}. 
Our starting point is the Hamiltonian
\begin{equation}
H=\sum_{k=1}^{N}\Omega_{k}|e\rangle\langle{g_{k}}|+h.c.,
\label{eq:HN}
\end{equation}
which can be seen as a
generalization of the $\Lambda$-system Hamiltonian and represents a system in which $N$ ground states
$|{g_{k}}\rangle$ are coupled to an excited level
$|e\rangle$. 
The level structure we have in mind is depicted in Fig. \ref{fig:levels}.
\noindent
The Hamiltonian (\ref{eq:HN})
admits $N-1$ dark states $|D_i\rangle$, i.e., zero-energy
eigenstates having no contribution from the excited state. 
The complex couplings $\Omega_k$ represent our control parameters.
Once they are fixed, the coding space, i.e., the $(N-1)$-fold degenerate eigenspace spanned by the dark
states, is  the orthocomplement of the vector ${\cal
N}^{-1/2}\sum_{k=1}^N \Omega_k |g_k\rangle \;({\cal
N}=\sum_{k=1}^N |\Omega_k|^2)$.
By setting $\Omega_1=\Omega_2=\ldots=\Omega_{N-1}=0$, we have that the
coding space $\cal C$ is spanned by the first $N-1$ ground
states.
We write the coupling constants in generalized spherical
coordinates
\begin{eqnarray}
\Omega_1&=&|\Omega|\sin(\theta_{1}), \nonumber \\
\Omega_2&=&|\Omega|e^{-i\phi_{2}}\cos(\theta_{1})\sin(\theta_{2}),\nonumber \\
\vdots 
\label{eq:cc} \\
\Omega_N&=&|\Omega|e^{-i\phi_{N}}\cos(\theta_{1})
\ldots\cos(\theta_{N-2})\sin(\theta_{N-1}).\nonumber
\label{eq:cc}
\end{eqnarray}
By explicitly computing the connection form $A$ and its
curvature for this system we have checked that it allows for
universal QC over $\cal C$ for any $N$. In particular, if $N=2^n+1$ $n$ qubits can be encoded in $\cal C$.
In the next two subsections we turn on the cases $N=3$ and $N=5$, or equivalently $n=1$
qubit and $n=2$ qubits. We will show how to realize single-qubit rotations -- thus, up to a phase factor, any
single-qubit gate -- and a 2-qubit phase-gate \cite{QC}. 

\subsection{(3+1)-level system}

One can realize a single-qubit encoding using a system
described by the Hamiltonian (\ref{eq:HN}) where $N=3$. As
computational basis we choose the first two ground states, i.e., we assign logical
values through the identities 
\begin{equation}
|g_{1}\rangle=|0\rangle, \; |g_{2}\rangle=|1\rangle,
\end{equation}
while the third ground state, $|g_{3}\rangle$, plays the role of an auxiliary
state, which
is necessary to achieve every single-qubit gate. 
It is well known that any single-qubit gate 
can be decomposed (up to a phase factor) in the product of three rotations, for instance
a rotation about the $z$ axis, one about the $y$ axis and one again about the $z$
axis, i.e., if $U$ is the single-qubit gate we want to build up, there exist three numbers, $\alpha,\beta,\gamma$,
such that $U=R_z(\gamma)R_y(\beta)R_z(\alpha)$, the Euler angles. 
Within our model Hamiltonian a rotation
about the $y$ axis can be obtained by putting the relative phases in Eq.
(\ref{eq:cc}) (with $N=3$)
$\phi_{i}=0$, and by adiabatically changing the amplitudes
$\theta_1,\theta_{2}$. In this case the connection is just
$A=A_{\theta_{2}}d\theta_{2}=-i\sin(\theta_{1})\sigma_{y}d\theta_{2}$,
where $\sigma_{y}$ is the $y$-Pauli matrix. We obtain the unitary operator (see App. \ref{app:sqg})
\begin{equation}
R_y(\beta)=\exp(i\beta\sigma_y),
\end{equation}
after a cycle $\chi$ in the ($\theta_1,\theta_{2}$)-submanifold, where the angle $\beta$ is given by
$\beta=\int_{S(\chi)}\cos(\theta_1)d\theta_1d\theta_2$ and $S(\chi)$ being the surface
enclosed by the loop $\chi$ on the ${\theta_1,\theta_2}$ submanifold. Up to a global phase, a rotation
about the $z$ axis is equivalent to the operator $\exp(i\alpha|1\rangle\langle 1|)$, which
is easily obtained by putting $\theta_1=\phi_3=0$ and by adiabatically performing a closed 
path $\chi'$ in the submanifold of ${\theta_2,\phi_2}$. The connection is 
 $A=A_{\phi_{2}}d\phi_{2}=-i\sin(\theta_{2})\alpha|1\rangle\langle 1|d\phi_{2}$. After a
 cycle we obtain
\begin{equation}
 \exp(i\alpha|1\rangle\langle 1|)=e^{i\alpha/2}R_z(\alpha),
 \label{eq:zrot}
\end{equation}
where $\alpha=\int_{S(\chi')}\sin(2\theta_2)d\theta_2d\phi_2$.

\subsection{(5+1)-level system}\label{sec:2qbit}
We assume two qubits mapped to a four level system.
Thus, to implement a two-qubit gate, we need $N=5$. We want to show how
to realize a phase gate, which assigns a phase only to one out of
the four computational basis states. As the computational basis
we choose $|D_i\rangle\equiv|g_{i}\rangle$, $i\leq 4$, whereby
the corresponding coupling constants $\Omega_{k}$ are initially
set to zero. The logical states can be identified as 
\begin{equation}
|g_{1}\rangle=|00\rangle,\;|g_{2}\rangle=|01\rangle,\;|g_{3}\rangle=|10\rangle,\;|g_{4}\rangle=|11\rangle, 
\label{eq:iden4}
\end{equation}
the fifth state playing the role of an ancilla.
Considering a closed path $\chi''$ in the
two-dimensional sub-manifold of the parameter space with
coordinates $(\theta_{4},\phi_{5})$ --  the other parameters
being kept to zero -- the connection is reduced to the simple
form:
$A=A_{\phi_{5}}d\phi_{5}=-i\sin^2(\theta_{4})|g_4\rangle\langle{g_4}|d\phi_{5}$.
This gives rise to the holonomy (see App. \ref{app:tqg})
\begin{equation}
\Gamma_{A}(\chi'')=\exp(i\alpha|{g_4}\rangle\langle{g_4}|)
\end{equation}
which, according to Eq. (\ref{eq:iden4}), precisely represents a 2-qubit phase gate, with the phase $\alpha$
given by $\int_{S(\chi'')}\sin(2\theta_{4})d\theta_{4}d\phi_{5}$.

\section{Physical realization}

\begin{figure}
\centerline{\epsfig{file=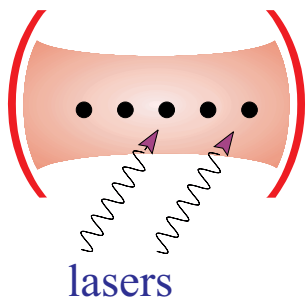,width=4cm}} 
\caption{Schematic representation of the setup.}
\label{fig:CQED}
\end{figure}

In the remaining part of the paper we discuss a possible physical realization of the -- up
to now quite abstract, but also very general -- concepts we introduced in the previuos
sections. Our proposal is based on atoms trapped inside an optical resonator 
(Fig. \ref{fig:CQED}). The atoms, which represent our qubits, interact individually with
laser beams and with a single quantized mode of the optical cavity.
The manipulation of the qubits (in terms of single- and 2-qubit gates)
involves only the laser beams, while the cavity mode is used
to prepare the system every time a 2-qubit gate is required.
Indeed, as we will see, we need to encode the information of two qubits in 
a single many-level atom and this is performed with the aid of the cavity
mode, to which all atoms are coupled. 

\subsubsection{single-qubit gate implementation}
The single-qubit gates
are easily implemented if we consider $(3+1)$-level atoms, i.e., atoms with
three ground states coupled by lasers to a single excited state. The Hamiltonian of the
single atom can be reduced to Eq. (\ref{eq:HN}) with $N=3$, where the coupling
constants  $\Omega_k, k=1,2,3$ are the Rabi frequencies of the lasers.  
Therefore for the feasibility of single-qubit gates the addressing of 
single levels in single atoms is required. 

\subsubsection{2-qubit gate implementation}

Once we have defined qubits, we need a way to couple them in a 
suitable way to make the computation universal.
In our case we are able to implement any 2-qubit transformation 
in a single atom, having $N=5$ ground (or meta-stable) states which can be coupled
by tunable lasers to a single excited one \cite{scalability}. Thus we have to deal with a
system in which the interacting part is described by the Hamiltonian 
Eq. (\ref{eq:HN}) with
$N=5$ and such that we can use the prescription of sec. \ref{sec:2qbit}. 
By adiabatically acting on the coupling lasers, which address the single levels of the
atom, any 2-qubit operation is achievable. \\
The complete picture is based on $(5+1)$-level atoms, 
single-qubit information being stored in the single atoms and 
single-qubit operations being performed in any single atom as discussed
above, using 3 out of the 5 ground states.
Let us consider two $(5+1)$-level atoms and let $|\alpha\rangle_1$ and $|\beta\rangle_2$ 
be the (logical) states of the first and the second atom, respectively.
A 2-qubit gate is performed in three steps:
\begin{enumerate}
 \item the two-qubit information is stored in the second system by the transfer 
   \begin{equation}
    |\alpha\rangle_1|\beta\rangle_2\to|0\rangle_1|\xi\rangle_2,
    \label{eq:inftran}
   \end{equation}
  where $\alpha$ ($\beta$) represents the first (second) digit of $\xi$ in binary notation;
 \item since we suppose that any $(5+1)$-level atom can be driven by the Hamiltonian 
      Eq. (\ref{eq:HN}), it is possible to obtain any (2-qubit) gate by manipulating the
      coupling constants, physically the Rabi frequencies $\Omega_k$.
      We have shown above how to obtain the phase gate 
      $U_{\alpha}=\exp(i\alpha|{11}\rangle\langle{11}|)$. 
      In this case it is sufficient to act only on two of the couplings
      $\Omega_k$, the others being turned off;
 \item  after the holonomic 2-qubit gate operation, the inverse transformation of Eq. (\ref{eq:inftran}) 
 is performed, and each qubit is encoded back in one of the two atoms.
\end{enumerate}
To pursue our purpose what is missing is a method to perform the information transfer 
Eq. (\ref{eq:inftran}). In order to be consistent with the holonomic paradigm, the information transfer has
to be adiabatically performed. 

\subsection{Information Transfer} 

\begin{figure}
\centerline{\epsfig{file=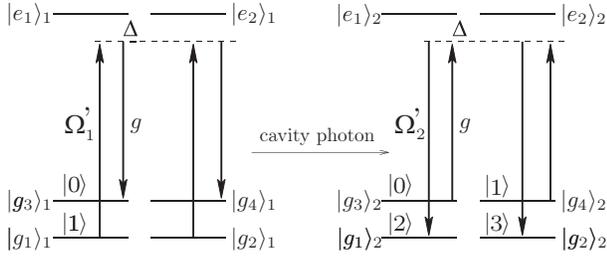,width=8cm}} \caption{Level
scheme and coupling laser lights for the information transfer.
The logical values are also shown (see the text).}
\label{fig:transfer}
\end{figure}

We will suggest two possible approaches to achieve the information transfer process. They involve, 
besides the ground states, excited states as well as
the single cavity mode. Thus, such processes will be affected by both spontaneous emission 
from the excited levels and imperfections of the cavity.
The first approach, that from now on we will call the optical scheme,
was envisaged in \cite{PellGardCZ}. It is based on an adiabatic transfer that
leaves almost unpopulated the excited levels,
thus reducing the influence of spontaneous emission. 
The second one, proposed here for the first time, is based on an adiabatic transfer that leaves
almost unpopulated 
the cavity mode, thus reducing the influence of cavity losses.   
We will call it  the motional scheme. We turn now to study in 
detail these two different approaches.\\
\indent
Both of them rely on the level scheme shown in Fig. \ref{fig:transfer}.
A laser addressing atom $i$ ($i=1,2$) with
Rabi frequency ${\Omega'}_i(t)$ couples $|g_1\rangle_i$ to the
excited state $|e_1\rangle_i$ and $|g_2\rangle_i$ to
$|e_2\rangle_i$, while $|g_3\rangle_i$ and $|g_4\rangle_i$ are
coupled by the cavity mode to, respectively, $|e_1\rangle_i$ and
$|e_2\rangle_i$.
\noindent
Note that the information transfer
is equivalent to state-swapping between the two atoms, provided the physical \
encoding is the one shown in Fig. \ref{fig:transfer}, i.e.,
$|g_3\rangle_1=|0\rangle_1$, $|g_1\rangle_1=|1\rangle_1$ for the first atom
and the 2-bit words, which we want to encode in the second
atom, are physically represented as  $|g_3\rangle_2=|0\rangle_2$,
$|g_4\rangle_2=|1\rangle_2$, $|g_1\rangle_2=|2\rangle_2$,
$|g_2\rangle_2=|3\rangle_2$. We have, for the logical state: 
\begin{eqnarray}
|g_3\rangle_1|g_3\rangle_2 \rightarrow |g_3\rangle_1|g_3\rangle_2 
\Leftrightarrow |0\rangle_1|0\rangle_2 \rightarrow |0\rangle_1|0\rangle_2
\nonumber\\ 
|g_3\rangle_1|g_4\rangle_2 \rightarrow |g_3\rangle_1|g_4\rangle_2 
\Leftrightarrow |0\rangle_1|1\rangle_2 \rightarrow |0\rangle_1|1\rangle_2 \nonumber\\ 
|g_1\rangle_1|g_3\rangle_2 \rightarrow |g_3\rangle_1|g_1\rangle_2  
\Leftrightarrow |1\rangle_1|0\rangle_2 \rightarrow |0\rangle_1|2\rangle_2 \nonumber\\
|g_1\rangle_1|g_4\rangle_2 \rightarrow |g_3\rangle_1|g_2\rangle_2 
\Leftrightarrow |1\rangle_1|1\rangle_2 \rightarrow |0\rangle_1|3\rangle_2 \nonumber
\end{eqnarray} 
Note that we used a different identification
of the logical state with respect to sec. \ref{sec:2qbit}. \\ 
To describe the state-transfer process, we will focus on the
evolution of one out of the two three-level systems which are
contained in each atom, i.e., the one formed by $|g_1\rangle_i$,
$|g_3\rangle_i$ and $|e_1\rangle_i$. \\
\indent
The evolution of the system, in presence of decoherence, will be described by a non-Hermitian Hamiltonian
in the framework of the quantum-jump approach to dissipative processes \cite{qjump}.\\
In the optical approach the following single-atom Hamiltonian is considered:
\begin{eqnarray}
H_{\rm opt}^i&=&\hbar[\omega_{e_1}|e_1\rangle_i\langle{e_1}|+\omega_{g_1}|g_1\rangle_i\langle{g_1}|+\omega_{g_3}|g_3\rangle_i\langle{g_3}|
+\omega_cb^{\dagger}b\nonumber\\
&+&\left({\Omega'}_i(t)e^{-i\omega_Lt}|e_1\rangle_i\langle{g_1}|+gb|e_1\rangle_i\langle{g_3}|+h.c.\right)]\nonumber\\
&-&i\gamma|{e_1}\rangle\langle{e_1}|-i\kappa{b^{\dagger}b}
\label{eq:Hoptcomp}
\end{eqnarray}
where $\omega_{e_1}$, $\omega_{g_1}$ and $\omega_{g_3}$ are the energies of the
state $|e_1\rangle_i$, $|g_1\rangle_i$ and $|g_3\rangle_i$, respectively, 
$g$ is the dipole-coupling constant between the cavity mode
and the atom and $b$ is the annihilation operator for the cavity
mode. As decoherence mechanism
we have considered the spontaneous emission $\gamma$ of the excited
levels $|e_j\rangle_i$ and the cavity loss rate $\kappa$. 
We will see that the transfer process is based on a dark state of the compound system 
$\cal{H}_{\rm atom1}\otimes\cal{H}_{\rm atom2}\otimes\cal{H}_{\rm cavity}$ which does not
involve any excited states and thus, within the adiabatic approximation,  
the main dissipative channel is due to the cavity loss rate. \\
In the motional approach we assume that the atoms are individually trapped in harmonic potentials.
The single-atom Hamiltonian contains also the harmonic trapping potential terms 
(see for instance \cite{Lin,Parkimble}):
\begin{eqnarray}
H_{\rm ext}^i&=&\hbar[\omega_{e_1}|e_1\rangle_i\langle{e_1}|+\omega_{g_3}|g_3\rangle_i\langle{g_3}|
+\omega_cb^{\dagger}b+{\nu}a^{\dagger}a \nonumber\\
&+&({\Omega'}_i(t)e^{-i\omega_Lt}|e_1\rangle_i\langle{g_3}|+h.c.)\nonumber\\
&+&g\sin(kx)(b|e_1\rangle_i\langle{g_3}|+h.c.)]\nonumber\\
&-&i\gamma|{e_1}\rangle\langle{e_1}|-i\kappa{b^{\dagger}b}
\label{eq:Hextcomp}
\end{eqnarray}
where $a \; (a^{\dagger})$ is the annihilation
(creation) operator for the harmonic motion, while the
sine function describes the standing-wave structure of the cavity field, with $k$
the wave number of the field and $x=[\hbar/(2m\nu)]^{1/2}(a^{\dagger}+a)$.
We will see that, under certain conditions, it is possible to obtain an effective Hamiltonian
which involves only the cavity mode and the harmonic motion. With such Hamiltonian
the transfer is based on a dark state with respect to the cavity and thus one
expects that the most important dissipative channel will be the spontaneuos 
decay. \\
First of all we will show, neglecting any dissipative mechanism, i.e.,
$\gamma=\kappa=0$, that indeed we can obtain the state-tranfer by acting in an adiabatic
fashion on the parameters of the Hamiltonians Eq. (\ref{eq:Hoptcomp}) and
Eq. (\ref{eq:Hextcomp}). Later on we will carry out in detail the analysis of the
effects of the dissipative channels.
In the next two subsections we do not consider any decay mechanism.

\subsubsection{Optical state transfer}

In the optical approach the starting point is the single-atom Hamiltonian
Eq. (\ref{eq:Hoptcomp}) (where for the moment $\gamma,\;\kappa = 0$).
Considering the energy $\hbar\omega_{g_1}$ as the zero of the
energy scale and transforming to the reference frame described
by the Hermitian operator
$A=\omega_cb^{\dagger}b+\omega_L|e_1\rangle\langle{e_1}|+(\omega_L-\omega_c)|g_3\rangle\langle{g_3}|$,
i.e., $|\psi\rangle\rightarrow{e^{itA}|\psi\rangle}$, one gets the
Hamiltonian \cite{Pell97}
\begin{equation}
H_{\rm opt}^i=\hbar[-\Delta|e_1\rangle_i\langle{e_1}|
+\left({\Omega'}_i(t)|e_1\rangle_i\langle{g_1}|+gb|e_1\rangle_i\langle{g_3}|+h.c.\right)].
\label{eq:Hopt1}
\end{equation}
In writing Eq. (\ref{eq:Hopt1}) we have considered the resonance
condition $\Delta_r=(\omega_{g_3}-\omega_{g_1})-(\omega_L-\omega_c)=0$
\cite{norescond} and we introduced the detuning
$\Delta=\omega_L-(\omega_{e_1}-\omega_{g_1})$ between the laser
frequency $\omega_L$ and the transition
$|g_1\rangle\rightarrow|e_1\rangle$. The 2-atom Hamiltonian
$H_{\rm opt}=H_{\rm opt}^1+H_{\rm opt}^2$ admits the eigenstate
\begin{eqnarray}
|\Psi\rangle_{\rm opt}&\propto&{\Omega'}_2g|g_1\rangle_1|g_3\rangle_2|0\rangle_{\rm cav}+
{\Omega'}_1g|g_3\rangle_1|g_1\rangle_2|0\rangle_{\rm cav}\nonumber\\
&-&{\Omega'}_1{\Omega'}_2|g_3\rangle_1|g_3\rangle_2|1\rangle_{\rm cav}.
\label{eq:transopt}
\end{eqnarray}
Thus by adiabatically changing the Rabi frequencies ${\Omega'}_1$
and ${\Omega'}_2$ applying a ``counterintuitive'' (\cite{RMPColl} and reference
therein) pulse (whereby
the pulse on atom 2 precedes that on atom 1) it is possible to
pass from the state $|g_3\rangle_1|g_1\rangle_2|0\rangle_{\rm cav}$
to the state $|g_1\rangle_1|g_3\rangle_2|0\rangle_{\rm cav}$. The
state-swapping is thus realized. 
Note that in such a scheme (see
Eq. (\ref{eq:transopt})) during the information transfer the
1-photon cavity state is populated.


\subsubsection{Motional state transfer}

\begin{figure}
\centerline{\epsfig{file=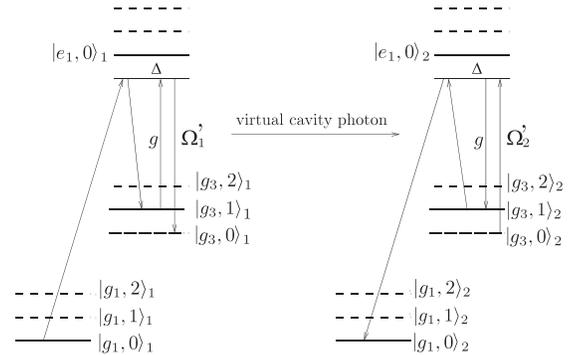,width=9cm}}
\caption{Example of information transfer by using external
degrees of freedom: $|g_1,g_3\rangle\rightarrow|g_3,g_1\rangle$,
photons, within the validity of Eq. (\ref{eq:Hexsm}), are never present in the cavity; the corresponding
logical operation is $|1\rangle_1
|0\rangle_2\rightarrow|0\rangle_1 |2\rangle_2$.}
\label{fig:transext}
\end{figure}

The single-atom Hamiltonian will be in this case the Eq. (\ref{eq:Hextcomp}) with 
$\gamma=\kappa=0$.
Considering the Lamb-Dicke limit, i.e., the size of the harmonic trap small compared with the optical wave-length, 
$k[\hbar/(2m\nu)]^{1/2}\ll{1}$,  and writing the
Hamiltonian in the reference frame given by the Hermitian operator (the zero-energy being $\omega_{g_3}$)
$A=\omega_L|e_1\rangle\langle{e_1}|+\omega_cb^{\dagger}b+{\nu}a^{\dagger}a$ one gets
\begin{eqnarray}
H_{\rm ext}^i&=&\hbar[-\Delta|e_1\rangle_i\langle{e_1}|+\left({\Omega'}_i(t)|e_1\rangle_i\langle{g_3}|+h.c.\right)\nonumber\\
&+&\eta(a^{\dagger}b|e_1\rangle_i\langle{g_3}|+abe^{-i2\nu{t}}|e_1\rangle_i\langle{g_3}|+h.c.)]
\label{eq:Hext1}
\end{eqnarray}
where $\eta=k[\hbar/(2m\nu)]^{1/2}$ is the Lamb-Dicke parameter.
It has been shown in \cite{Lin,Parkimble} that under the
conditions $\Delta\gg{|{\Omega'}_i|},g\eta$ a Hamiltonian which
does not involve the atomic internal degrees of freedom is
obtainable. By adiabatic elimination and using the rotating wave
approximation (RWA) we find the effective Hamiltonian (see App. \ref{app:opteffH})
\begin{equation}
H_{\rm tr}=\hbar\sum_{i=1}^{2}G_{i}(t)a_{i}^{\dagger}b+h.c.,
\label{eq:Hexsm}
\end{equation}
where we introduced the coupling parameters
$G_i=-g\eta{\Omega'}_i(t)/\Delta$. The number of excitations
$\sum_{i=1}^2a_i^{\dagger}a_i+b^{\dagger}b$ is a conserved
quantity, in particular the zero-excitation eigen-space is
spanned by the vacuum state $|0\rangle_1^{\rm cm}
|0\rangle_2^{\rm cm}|0\rangle_{\rm cav}$ and the 1-excitation
eigen-space by the three states $\{|1\rangle_1^{\rm cm}|0\rangle_2^{\rm
cm}|0\rangle_{\rm cav},|0\rangle_1^{\rm cm}|1\rangle_2^{\rm
cm}|0\rangle_{\rm cav},|0\rangle_1^{\rm cm}|0\rangle_2^{\rm
cm}|1\rangle_{\rm cav}\}$, where $|n\rangle_i^{\rm cm}$ is the eigenstate
of the free external Hamiltonian of the $i$th atom, it satisfies
$a_i^{\dagger}a_i|n\rangle_i^{\rm cm}=n|n\rangle_i^{\rm cm}$.
Inside the last sub-space there exists the
dark state (dark with respect to the cavity mode)
\begin{eqnarray}
|\Psi\rangle_{\rm ext}&\propto&{\Omega'}_2|1\rangle_1^{\rm
cm}|0\rangle_2^{\rm cm}|0\rangle_{\rm cav}- {\Omega'}_1|0\rangle_1^{\rm
cm}|1\rangle_2^{\rm cm}|0\rangle_{\rm cav}. \label{eq:transtr}
\end{eqnarray}
In an analogous way to what described in the previous Section, by
adiabatically changing the Rabi frequencies ${\Omega'}_i(t)$ it is
possible to pass from the state $|1\rangle_1^{\rm
cm}|0\rangle_2^{\rm cm}|0\rangle_{\rm cav}$ to the state
$|0\rangle_1^{\rm cm}|1\rangle_2^{\rm cm}|0\rangle_{\rm cav}$ without
populating the cavity mode which, in the case of a nonzero cavity
loss rate $\kappa$, is a source of decoherence. \\
We are now ready to describe how the state-swapping can be performed in three steps
starting without photons in the cavity and
zero-motional-excitation, see Fig. \ref{fig:transext}: first of
all the logical state of qubit 1 is swapped onto the motional
state of atom 1 by adiabatic passage via a dark state; then we
utilize the dark state (\ref{eq:transtr}) to transfer the
motional state to the second atom and this is swapped onto the
internal state (of the second atom) by the same adiabatic pulse
sequence as in first step, ending with a global state without any
excitation, i.e., no photons, no quanta of harmonic motion.

\section{Decoherence effects}

We will now come back to study what are the limitations imposed by the 
dissipative channels on the transfer process.
As stated well before we want to take into account excited level spontaneous 
emission and cavity loss. Thus the evolution of the system is 
described by the Eq. (\ref{eq:Hoptcomp}) and Eq. (\ref{eq:Hextcomp}),
where the presence of the spontaneous emission rate $\gamma$ and the cavity loss rate 
$\kappa$ give rise to a non-hermitian evolution. 

\subsection{Optical state transfer}

The optical state transfer is based on the dark state Eq. (\ref{eq:transopt}).
First of all we note that it has a non-zero projection onto the 1-photon cavity state
and so a lossy cavity will tend to destroy such a state. The dark state is, indeed, 
not an eigenstate of the Hamiltonian Eq. (\ref{eq:Hoptcomp}) for $\kappa\neq 0$.
In order for the influence of the cavity loss rate to be small, the condition 
$\kappa\int_0^T P_{\rm 1-ph}(t)dt\ll 1$ has to be satisfied, where 
$P_{\rm 1-ph}(t)=\langle\Psi_{\rm opt}|b^\dagger b|\Psi_{\rm opt}\rangle$ is the population of the
cavity mode during the adiabatic evolution and $T$ is the time process.
Since the integral in the inequality is
always smaller than $T\max_t\{P_{\rm 1-ph}(t)\}$, inserting into the expression 
Eq. (\ref{eq:transopt}) Gaussian-shaped laser pulses with peak value
$\Omega$, time separation $aT$ and variance $\tau T/\sqrt{2}$, one
can give the following requirement 
\begin{equation}
\kappa T {\rm max}_t\{P_{\rm 1-ph}(t)\}=\kappa T
\left(1+\frac{2g^2}{\Omega^2}e^{2a^2/\tau^2}\right)^{-1} \ll 1.
\label{eq:kopt}
\end{equation}

Furthermore, in any real process (finite time, finite energy), the state of the system will
precess around the dark state, instead of following it in a perfect adiabatic way. This
means that some population reaches the leaky states $|e\rangle_i$. 
One should calculate the population $P_e$ 
of such unwanted states during the adiabatic process and impose the 
condition $\gamma \int_0^T P_e(t)dt \ll 1$. 
We give here and in what follows some simplified conditions. 
The condition for adiabatic passage can be stated as 
\begin{equation}
\tilde\Omega T \gg 1.
\end{equation} 
Here $\hbar \tilde{\Omega}$ is a parameter expressing an estimate 
of the global value of the differnce between the dark-state energy (zero) and 
the smallest (non-zero) eigenenergy, $E_{\rm min}(t)$, of the Hamiltonian 
Eq. (\ref{eq:Hopt1}) during the transfer process. 
The condition on $\gamma$ reads:
\begin{equation}
\frac{\gamma T}{(\tilde\Omega T)^2}\ll 1.
\end{equation}
The transfer time, $T$, is bounded from below and from above.
Let us compare the on-resonance case, i.e., $\Delta = 0$ and the far off-resonance case,
i.e., $\Delta \gg |{\Omega'}_i|, g$. We evaluate $\tilde{\Omega}$ as a time average of $E_{\rm min}(t)$ 
\cite{RMPColl}.
In these regimes we find 
\begin{eqnarray}
\tilde\Omega_{\Delta=0}&=&\left[g^2+\frac{1}{2}\left(\Omega_{\rm eff}^2-
\sqrt{({\Omega'}_1^2-{\Omega'}_2^2)^2+4g^4}\right)\right]^{1/2}, \\
\tilde\Omega_{\text{large $\Delta$}}&=&\frac{1}{\Delta}\left[g^2+\frac{1}{2}\left(\Omega_{\rm
eff}^2-\sqrt{({\Omega'}_1^2-{\Omega'}_2^2)^2+4g^4}\right)\right],
\end{eqnarray} 
with $\Omega_{\rm eff}^2={\Omega'}_1^2+{\Omega'}_2^2$. \\
Since we know that the adiabatic passage works better when
$a \simeq \tau$ \cite{RMPColl}, from Eq. (\ref{eq:kopt}), we learn that it is favourable to have $g \gg
\Omega$. The tranfer time is restricted to be:
\begin{equation}
\frac{2g^2}{\kappa\Omega^2}\gg T \gg \frac{\gamma}{2\langle\Omega_{\rm eff}\rangle^2}C,
\end{equation}
where $C=1$ for $\Delta =0$ and $C=8\Delta^2/\langle\Omega_{\rm eff}^2\rangle$ for large detuning.
Provided that the previous inequality is
satisfied on both sides by a factor $\alpha$, we obtained that, at least, it must be
$\kappa\gamma\le(g/\alpha)^2$ for $\Delta=0$ and
$\kappa\gamma\le(g/\alpha)^2(\Omega/\Delta)^2$ for large $\Delta$. Then it is clear that the optical scheme
works better in the on-resonance regime.

\subsection{Motional state transfer}

For the transfer procedure via motional state swap, it is possible, as we have seen,
to obtain a Hamiltonian involving just the external degrees of
freedom and the cavity mode. This is true also in the presence of decoherence
processes, where one obtains the non-Hermitian Hamiltonian 
\begin{equation}
H_{\rm eff,i}^{RWA}=\frac{\eta{g}{\Omega'}_i}{\Delta+i\gamma}(a_ib^{\dagger}+a_i^{\dagger}b)-i\kappa{b^{\dagger}b}
-i\frac{\gamma{\Omega'}_i^2}{\Delta^2+\gamma^2}.
\label{eq:HeffRWA}
\end{equation}
The details of the calculation are given in  App. \ref{app:moteffH}, where also
the Hamiltonian without RWA is shown. To obtain Eq. (\ref{eq:HeffRWA}) 
the conditions $\Delta \gg |{\Omega'}_i|, \eta g$ are imposed.
If $T$ is the process time, the conditions we can
give on $\gamma$ are (first order pertubation theory):
\begin{equation}
 \gamma T\left(\frac{|{\Omega'}_i|}{\Delta}\right)^2, \;  
 \gamma T\left(\frac{\eta g}{\Delta}\right)^2 \ll 1.
\end{equation}
The motional scheme is based on a dark-state with respect to the cavity mode (see Eq.
(\ref{eq:transtr})).
The actual state of the system during the evolution will precess
around the dark state and so we will have a certain population $P_{\rm 1-ph}$ of the 1-photon 
cavity state. In order to give a condition on $\kappa$ we have to estimate such 
population along the adiabatic process.  
For Gaussian-shaped pulse we found (see App. \ref{app:pop}):
\begin{equation}
P_{\rm 1-ph}\leq \frac{a^2}{\tau^4}e^{a^2/\tau^2} \left( \frac{g\eta\Omega}{\Delta}
T\right)^{-2}.
\label{eq:pop1phmot}
\end{equation} 
The condition on $\kappa$ reads: $T\kappa P_{\rm 1-ph}\ll 1$.
Thus considering for convenience $m=\max\{\eta g,\; \Omega\}$, we see that the process time has to be
($a\simeq\tau$):
\begin{equation}
\frac{1}{\gamma}\left(\frac{\Delta}{m}\right)^2\gg T 
\gg \frac{1}{(\tau\eta)^2}\frac{\kappa}{g^2}\left(\frac{\Delta}{\Omega}\right)^2.
\end{equation} 

First of all note the inverted role of $\gamma$ and $\kappa$ in the optical and in the
motional scheme in restricting the process time. Then, if the inequality is satisfied on both side by a factor $\alpha$
we have that $\kappa\gamma\le(g/\alpha)^2(\tau\eta\Omega/m)^2$.
We see that, in the motional scheme, we have a much more strict
restriction on $\kappa\gamma$ due to the presence of the small Lamb-Dicke parameter.

\subsection{Modified optical scheme}

In this section we describe how to enrich the optical scheme to obtain a
new scheme, where the role of the decoherence mechanisms is the same as in the motional
scheme, without the Lamb-Dicke parameter. \\

In the case of large detuning , i.e., $\Delta \gg |{\Omega'}_i|, g$, we can perform an 
adiabatic elimination of the excited states. Starting from the Hamiltonian Eq. (\ref{eq:Hoptcomp}), we obtain the following 
2-level Hamiltonian (see App. \ref{app:opteffH} and \cite{Pell97}):
\begin{eqnarray}
H_{\rm eff,i}^{\rm opt}&=&\frac{{\Omega'}_i^2}{\Delta+i\gamma}|g_1\rangle_i\langle{g_1}|
+\frac{g^2}{\Delta+i\gamma}b^\dagger b|g_3\rangle_i\langle{g_3}| \nonumber \\
&+&\frac{g{\Omega'}_i}{\Delta+i\gamma}(b^\dagger |g_3\rangle_i\langle{g_1}|+b|g_1\rangle_i\langle{g_3}|)
-i\kappa{b^\dagger b}.
\label{eq:Heffopt}
\end{eqnarray}
The total Hamiltonian 
$H_{\rm eff}^{\rm opt}=H_{\rm eff,1}^{\rm opt}+H_{\rm eff,2}^{\rm opt}$, 
when $\kappa = 0$, admits, as expected, the zero-energy eigenstate Eq. (\ref{eq:transopt}). 
If it is possible to compensate the total Stark
shift of the ground state $|g_1\rangle_i$ -- i.e., the real part of the complex energy 
of  $|g_1\rangle_i$ in Eq. (\ref{eq:Heffopt}) --
one obtains a Hamiltonian which has a similar
structure to the one obtained for the motional scheme, Eq. (\ref{eq:Hexsm}):
\begin{eqnarray}
\tilde{H}_{\rm eff,i}^{\rm opt}&=&-2i\gamma\frac{{\Omega'}_i^2}{\Delta^2+\gamma^2}|g_1\rangle_i\langle{g_1}|
+\frac{g^2}{\Delta+i\gamma}b^\dagger b|g_3\rangle_i\langle{g_3}| \nonumber \\
&+&\frac{g{\Omega'}_i}{\Delta+i\gamma}(b^\dagger |g_3\rangle_i\langle{g_1}|+b|g_1\rangle_i\langle{g_3}|)
-i\kappa{b^\dagger b}.
\label{eq:Hstab}
\end{eqnarray}
Physically the compensation can be realized by coupling the state $|g_1\rangle$ with another
auxiliary state via a laser pulse which has the same Rabi frequency as the one that couples 
$|g_1\rangle_i$ to $|e_1\rangle_i$, i.e., ${\Omega'}_i$, and a detuning $-\Delta$. For simplicity, the
auxiliary state is supposed to have the same spontaneous emission rate $\gamma$ as 
$|e_1\rangle_i$ have. 
When $\gamma =0$ (while $\kappa$ can be non-zero), the Hamiltonian Eq. (\ref{eq:Hstab})
admits the dark state
with respect to the cavity mode 
\begin{equation}
|\Psi\rangle\propto \Omega'_2|g_1\rangle_1|g_3\rangle_2|0\rangle_{\rm cav}+
\Omega'_1|g_3\rangle_1|g_1\rangle_2|0\rangle_{\rm cav}.
\end{equation} 
To obtain the requirement on $\kappa$, we estimate the population of the 1-photon cavity 
state during the adiabatic transfer. 
Using the same notation we used above we obtained (see Appendix \ref{app:pop}):
\begin{equation}
P_{\rm 1-ph}\leq \frac{a^2}{\tau^4}e^{a^2/\tau^2} \left( \frac{g\Omega}{\Delta}
T\right)^{-2}.
\label{eq:pop1phopt}
\end{equation} 
The condition on $\kappa$ is $T\kappa P_{\rm 1-ph}\ll 1$. 
The presence of the auxiliary levels impose on $\gamma$ the restriction
\begin{equation}
\gamma T\left(\frac{|{\Omega'}_i|}{\Delta}\right)^2 \ll 1.
\end{equation}
Thus, for the process time, $T$, we have (as usual $a \simeq \tau$)
\begin{equation}
\frac{1}{\gamma}\left(\frac{\Delta}{\Omega}\right)^2\gg T 
\gg \frac{1}{\tau^2}\frac{\kappa}{g^2}\left(\frac{\Delta}{\Omega}\right)^2,
\end{equation}
which is the same condition we gave for the motional scheme, but without the small Lamb-Dicke
parameter. For instance the condition on $\gamma\kappa$, obtained as explained before,
reads $\gamma\kappa \le (g/\alpha)^2 \tau^2$. 

\subsection{Summary}

In summary, using the motional scheme it is possible to reduce the
effects of $\kappa$ by increasing the process time, since it is
based on the adiabatic transfer via a dark state with respect to
the cavity. The process cannot be too long in order to avoid
spontaneous emission effects, which are however reduced by
choosing a large detuning. On the other hand the optical scheme,
based on a dark state with respect to the excited levels, can
avoid the effect of the spontaneous emission by increasing the
process time. In this case, however, the process must be fast with
respect to the inverse of the cavity loss rate. Finally, provided
we can use a stabilizing pulse to compensate for the
time-dependent energy of the ground state (see also
\cite{Pell97}), the optical scheme in the large detuning regime
could be able to operate in the same fashion as the motional scheme
but faster (basically by a factor of the order of the Lamb-Dicke
parameter $\eta$) and then to allow for a larger spontaneuos emission rate $\gamma$. 

\subsection{Numerical results}

\begin{figure}
\centerline{\epsfig{file=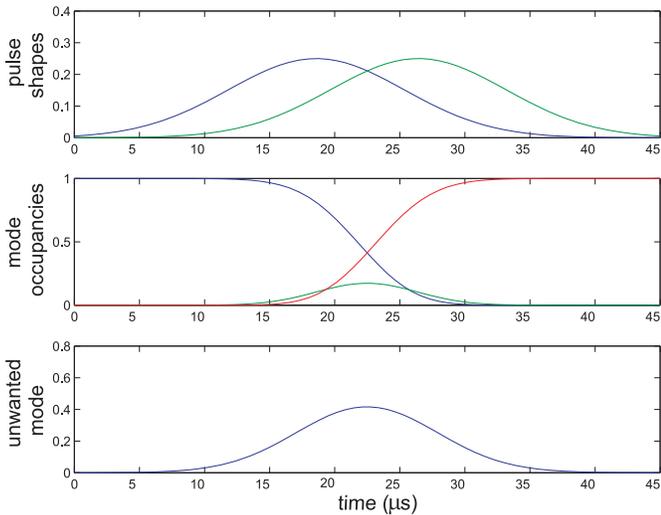,width=9cm}}
\caption{Adiabatic evolution of the system. The figure 
represents the evolution of the projection of the dark-state, for the optical scheme,
onto the initial basis states. We see that, essentially, 
only the projection onto the states involved in the dark-state are populated.}
\label{fig:extrans}
\end{figure}

\begin{figure}
\centerline{\epsfig{file=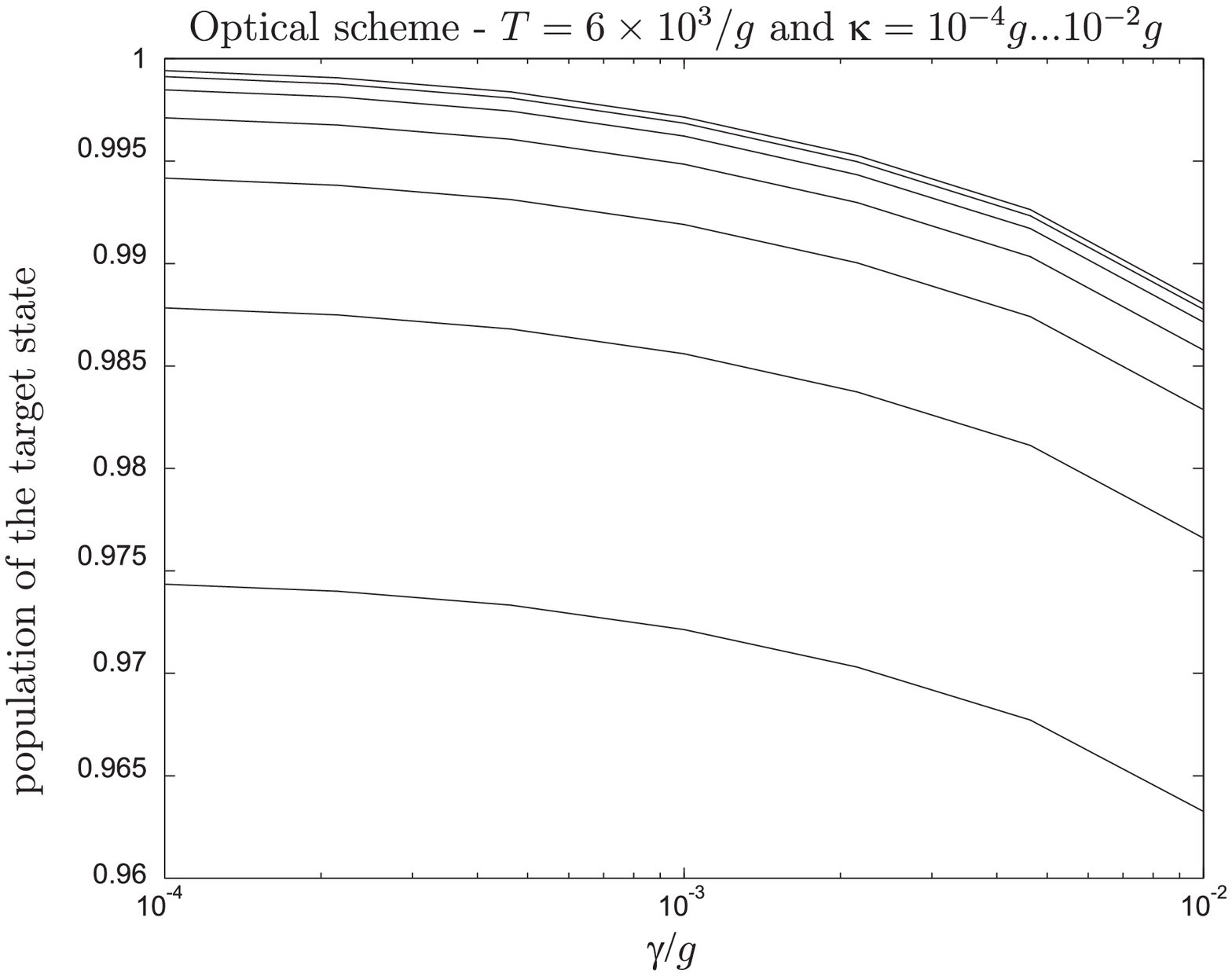,width=9cm}}
\centerline{\epsfig{file=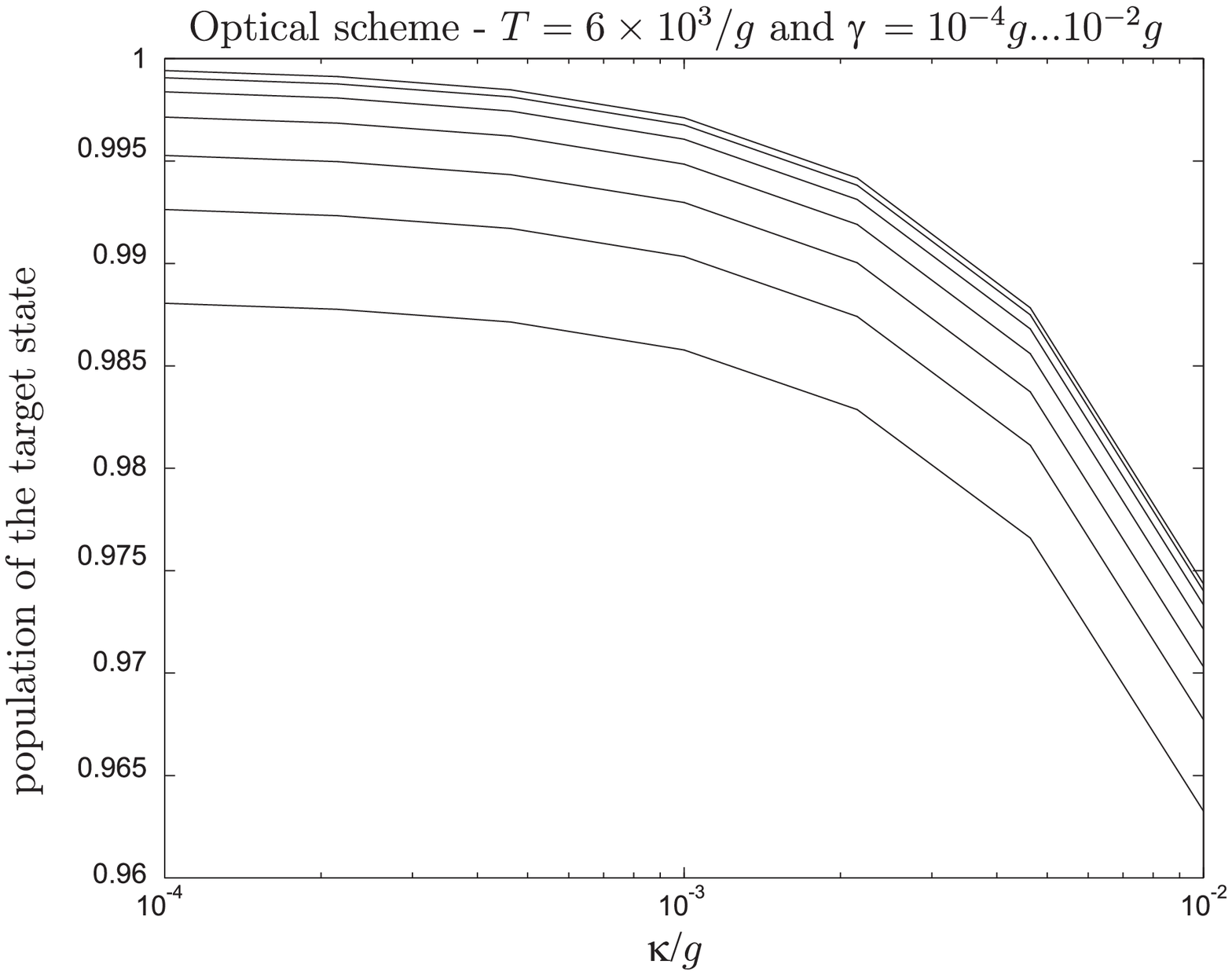,width=9cm}}
\caption{Population of the ``target'' state for the optical scheme as a function of the
spontaneous decay rate, $\gamma$, for various values of the cavity loss rate, $\kappa$ (up); 
as a function of the cavity loss rate, $\kappa$, for various values of the spontaneous 
decay rate, $\gamma$ (down).}
\label{fig:optdec}
\end{figure}

\begin{figure}
\centerline{\epsfig{file=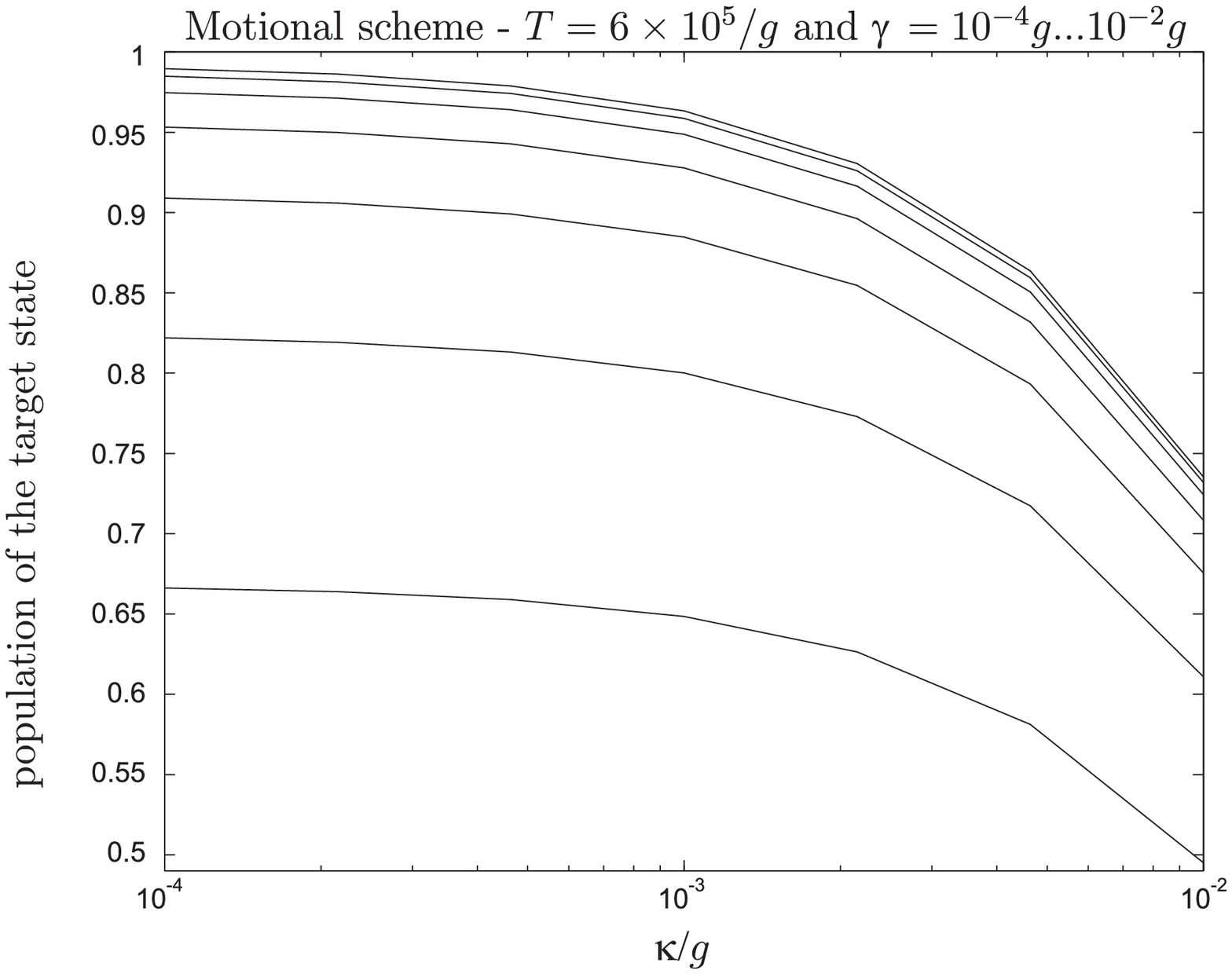,width=9cm}}
\centerline{\epsfig{file=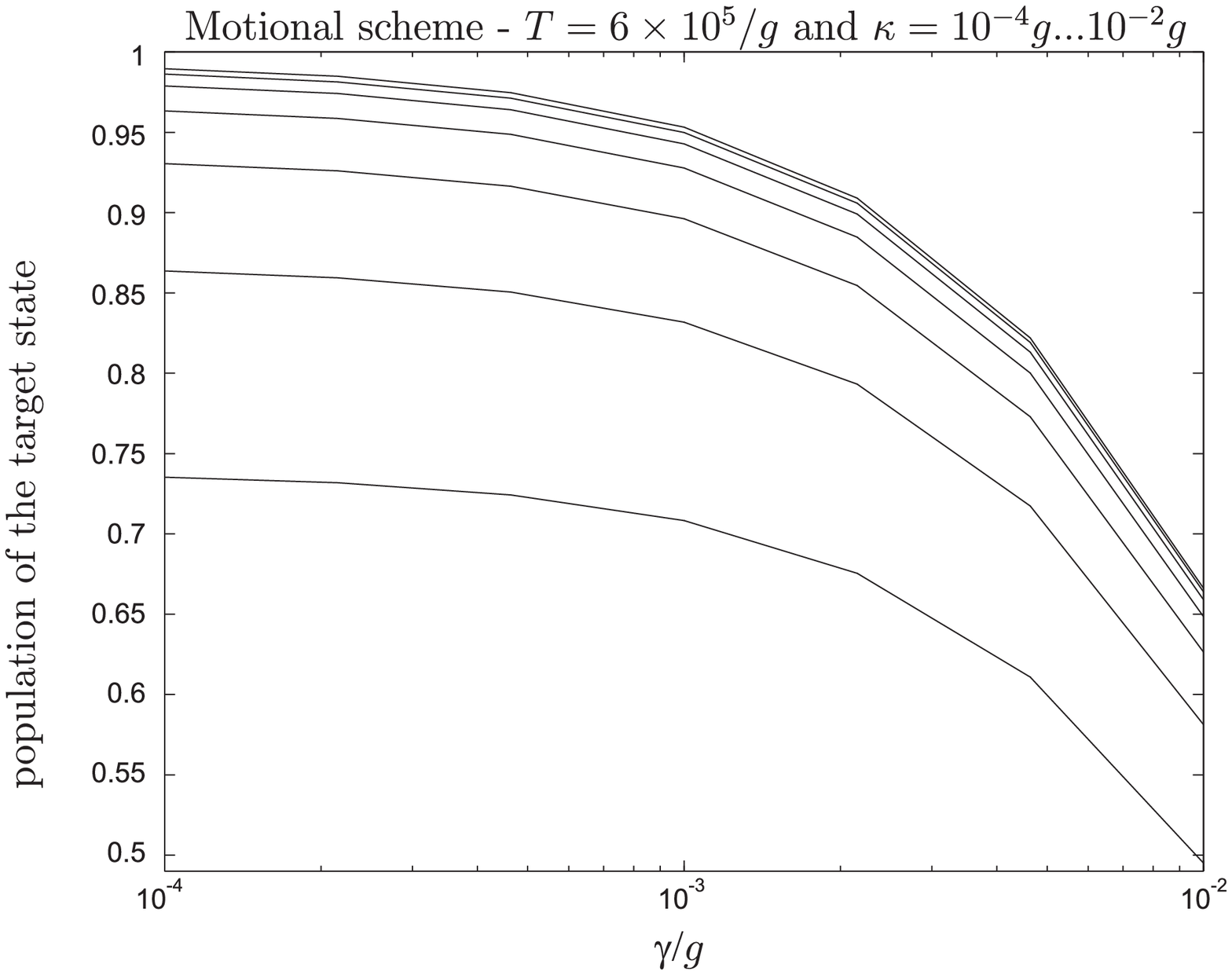,width=9cm}}
\caption{Population of the ``target'' state for the motional scheme as a function of the
spontaneous decay rate, $\gamma$, for various values of the cavity loss rate, $\kappa$ (up); 
as a function of the cavity loss rate, $\kappa$, for various values of the spontaneous 
decay rate, $\gamma$ (down).}
\label{fig:motdec}
\end{figure}

We performed numerical simulations for the suggested methods. They confirm what we have stated in the
last sections. We show in Fig. (\ref{fig:optdec}) and Fig. (\ref{fig:motdec}) the transfer fidelity,
respectively, for the optical scheme and for the motional one, under the same condition of the parameter
$\gamma$, $\kappa$, $\Omega$ and $g$. We expressed all the quantities in unity of the coupling, $g$,
between the atoms and the cavity mode. We simulated the evolution of the system as driven by Gaussian-shaped pulses. We
have followed the evoltion of the projections of the state onto the bare states, i.e., 
$|g_i\rangle$ and $|e_i\rangle$. We are interested in the population of the transfer's target state 
Fig. (\ref{fig:extrans}).  
It is this population that it is plotted, as a function of $\kappa$ and $\gamma$, in the 
Figs. (\ref{fig:optdec}) and (\ref{fig:motdec}), the time is fixed. In such shown figures we use: $\Omega=0.05g$, 
$\Delta=0$, for the optical scheme, $\Delta=10g$ and $\eta=0.1$ for the motional scheme. \\
  
As expected from the previously given inequalities the optical
transfer is more affected by the presence of the cavity loss, than by the presence of the spontaneous
emission, the vice versa is true for the motional scheme. In this parameter regime, the optical scheme
works much better by means of both fidelity and transfer time, than the motional one. However is worth
to remind the different role played by the two decoherence channels as it has been summarized above. In
principle for a bad cavity and a small spontaneous emission rate, one shall have, waiting enough time, a
fidelity unreacheable with the optical scheme. 

\section{Conclusion}

In this paper we have presented an implementation of quantum
computation based on holonomic operations, i.e., non-Abelian
geometric phases, with trapped neutral atoms. The method is a
generalization of a previous proposal exploiting ions
\cite{IgnaLu}. We showed explicitly how to realize a general
unitary transformation on $n$ qubits encoded in a $(2^n+2)$-level
atom. We also developed a scalable scheme, where each qubit is
stored in one atom. In this case, two-qubit operations are
performed in one of the two atoms, which encodes both atoms'
initial logical state after a properly designed transfer process.
We discussed in detail two possible procedures for such a state
transfer, relying on adiabatic passage via a dark state of the
two-atoms-plus-cavity system. The two schemes (the second of
whose is originally proposed here) differ in the atomic degrees
of freedom involved -- respectively, internal or external. We
discussed advantages and limitations of both schemes in the
presence of decoherence, for different parameter regimes, and we
found that the second proposal is more suitable for a situation
with a comparatively lossy cavity, as it might be the case, e.g.,
with atomic micro-traps coupled to surface-mounted micro-cavities
in the context of the so-called {\em Atom Chips} \cite{AtomChip}.

\acknowledgments

This research has been supported by the Austrian Science
Foundation, the Institute for Quantum Information GmbH, the
Istituto Trentino di Cultura and the European Commission through
contracts ERB-FMRX-CT96-0087, IST-1999-11055 (ACQUIRE) and
HPMF-CT-1999-00211.

\end{multicols}

\appendix

\section{Holonomies}

\subsection{Single-qubit gate}\label{app:sqg}

As described in the text any single-qubit gate can be realized by holonomic means if
we consider the Hamiltonian Eq. (\ref{eq:HN}) with $N=3$.
We write the coupling parameters using a kind of spherical coordinates:
\begin{eqnarray}
\Omega_{1}&=&|\Omega|\sin(\theta_{1}), \nonumber\\
\Omega_{2}&=&|\Omega|e^{-i\phi_{2}}\cos(\theta_{1})\sin(\theta_{2}),\nonumber\\
\Omega_{3}&=&|\Omega|e^{-i\phi_{3}}\cos(\theta_{1})\cos(\theta_{2}).
\end{eqnarray}
The Hamiltonian admits two zero-eigenvalue eigenvectors $|\psi^1\rangle$ and $|\psi^2\rangle$, which can be
written in terms of the ground-states $|g_k\rangle$, $k=1,2,3$ as
\begin{eqnarray}
|\psi^{1}\rangle&=&\cos(\theta_{1})|g_{1}\rangle-\sin(\theta_{1})[e^{-i\phi_{2}}\sin(\theta_{2})|g_{2}\rangle
+e^{-i\phi_{3}}\cos(\theta_{2})|g_{3}\rangle]\nonumber\\
|\psi^{2}\rangle&=&-e^{-i\phi_{2}}\cos(\theta_{2})|g_{2}\rangle+e^{-i\phi_{3}}\sin(\theta_{2})|g_{3}\rangle.
\end{eqnarray}
We fixed the two relative phases $\phi_2$ and $\phi_3$ to zero and calculated the connection components
$A_{\mu}^{ij}=\langle\psi^i|{\partial}/\partial\lambda_\mu|\psi^j\rangle$ where
$\{\lambda_\mu\}=\{\theta_k,\phi_j; k=1,2,3; j=2,3\}$. We obtained the
connection
\begin{equation}
A=A_{\theta_2}d\theta_2=-i\sin(\theta_1)\sigma_yd\theta_2,
\end{equation}
where $\sigma_y$ is the $y$-Pauli matrix.
The related unitary operation is $U(\chi)_A={\bf P}\exp \int_\chi A={\bf P}\exp \int_\chi
A_{\theta_2}d\theta_2$, where the integral is along a loop $\chi$ in the sub-manifold
${\theta_1,\theta_2}$. The line integral can be converted using the Stokes
theorem in the surface integral $\int_{S(\chi)}F_{\theta_1\theta_2}d\theta_1d\theta_2$, where
\begin{equation}
F_{\theta_1\theta_2}=\partial_{\theta_1}A_{\theta_2}-\partial_{\theta_2}A_{\theta_1}=
-i\cos(\theta_1)\sigma_y
\end{equation}
are the components of the so-called curvature 2-form, and
$S(\chi)$ is the surface enclosed by the loop $\chi$
 in the ($\theta_1,\theta_2$)-plane. The unitary operator takes the form:
\begin{equation}
U(\chi)=\exp{-i\sigma_y}\int_{S(\chi)}\cos(\theta_1)d\theta_1d\theta_2
\end{equation}
with $\sigma_y$ the $y$-Pauli matrix, i.e., we have obtained a qubit rotation around the $y$-axis of the Bloch
sphere. In the same way one can obtain the gate Eq. (\ref{eq:zrot}), fixing $\theta_1=\phi_3=0$.
In this case $|\psi^{1}\rangle=|g_1\rangle$ for any values of the parameters and so, 
after a cycle in the remaing parameter submanifold, only the 
state $|g_2\rangle$ acquires a phase. 
We stated also in the text that by choosing in a suitable way the loop in the manifold
$\{\theta_k,\phi_j; k=1,2,3; j=2,3\}$ it is possible to obtain any single-qubit gate.
Showing the feasibility of the 2-qubit phase gate is enough to conclude the universality of this approach. We
stress once more that in this approach, anyway, instead of thinking how a particular gate
is decomposed in single- and two-qubit gates and then to try to realize them, can be much easier from the
experimental point of view to search suitable loops (experimentally it means to search for the simplest loops) 
to perform the quantum gate we want. Indeed we have that, given any unitary operator, $U$, there exists a closed 
path, in the parameter space such that the holomony it generates coincide with $U$.

\subsection{Two-qubit gate}\label{app:tqg}

Realizing the phase-gate does not require more effort than realizing a single-qubit rotation.
In this case the Hamiltonian is given by Eq. (\ref{eq:HN}) with $N=5$.
We write the coupling parameters using a kind of spherical coordinates:
\begin{eqnarray}
\Omega_1&=&|\Omega|\sin(\theta_1), \nonumber\\
\Omega_2&=&|\Omega|e^{-i\phi_2}\cos(\theta_1)\sin(\theta_2),\nonumber\\
\Omega_3&=&|\Omega|e^{-i\phi_3}\cos(\theta_1)\cos(\theta_2)\sin(\theta_2), \nonumber\\
\Omega_4&=&|\Omega|e^{-i\phi_4}\cos(\theta_1)\cos(\theta_2)\cos(\theta_3)\sin(\theta_4),\nonumber\\
\Omega_5&=&|\Omega|e^{-i\phi_5}\cos(\theta_1)\cos(\theta_2)\cos(\theta_3)\cos(\theta_4).
\end{eqnarray}
This Hamiltonian admits four zero-eigenvalue eigenvectors $|\psi^i\rangle$
($i=1,2,3,4$), which can be
written in terms of the ground states $|g_k\rangle$ ($k=1,2,3,4,5$) as
\begin{eqnarray}
|\psi^1\rangle&=&\cos(\theta_1)|g_1\rangle-\sin(\theta_1)\{e^{-i\phi_2}\sin(\theta_2)|g_2\rangle
+\cos(\theta_2)\left[e^{-i\phi_3}|g_3\rangle+\cos(\theta_3)
\left(e^{-i\phi_4}\sin(\theta_4)|g_4\rangle+e^{-i\phi_5}\cos(\theta_4)|g_5\rangle\right)\right]\}, \nonumber\\
|\psi^2\rangle&=&e^{-i\phi_2}\cos(\theta_2)|g_2\rangle-\sin(\theta_2)
\left[e^{-i\phi_3}\sin(\theta_3)|g_3\rangle+\cos(\theta_3)(e^{-i\phi_4}\sin(\theta_4)|g_4\rangle
+e^{-i\phi_5}\cos(\theta_4)|g_5\rangle\right],  \nonumber\\
|\psi^3\rangle&=&e^{-i\phi_3}\cos(\theta_3)|g_3\rangle-\sin(\theta_3)
(e^{-i\phi_4}\sin(\theta_4)|g_4\rangle+e^{-i\phi_5}\cos(\theta_4)|g_5\rangle), \nonumber\\
|\psi^{4}\rangle&=&-e^{-i\phi_{4}}\cos(\theta_{4})|g_{4}\rangle+e^{-i\phi_{5}}\sin(\theta_{4})|g_{5}\rangle.
\end{eqnarray}
It is worth noting that, when all the parameters (actually the angles
$\theta_i$) are fixed to zero, the previous
eigenstates coincide with the 4 ground states $|g_1\rangle,..,|g_4\rangle$: we will consider always paths which start
and end at such a point.
We fixed to zero the parameters $\theta_1,\theta_2,\theta_3$ -- i.e., the coupling constants
$\Omega_1,\Omega_2,\Omega_3$ -- and the relative phases $\phi_2,\phi_3,\phi_4$, thus the connection gets the
simple expression
\begin{equation}
A=A_{\phi_5}d\phi_5=-i\sin^2\theta_4|g_4\rangle\langle g_4|d\phi_5.
\end{equation}
It gives rise to the unitary operator $U(\chi'')_A={\bf P}\exp \int_\chi'' A_{\theta_2}d\theta_2$, which once again
can be easily calculated by using the Stokes theorem to convert to a surface integral the line integral.
Indeed, by introducing the curvature 2-form $F$, which in this case has just one non-zero component
$F_{\phi_5\theta_4}=i\sin(2\theta_4)$, one gets
\begin{equation}
U(\chi'')=\exp{-i\int_{S(\chi'')}\sin(2\theta_4)d\theta_4d\phi_5}|g_4\rangle\langle{g_4}|.
\end{equation}
The last expression is precisely a phase gate, i.e., an operation that assigns a
phase -- equal to the surface-integral --
to one ($|g_4\rangle$) out of four states.

\section{Effective Hamiltonians}

\subsection{Three-level atom in a cavity: Effective Hamiltonian}\label{app:opteffH}

In this section we are going to find out an effective, approximate Hamiltonian, starting from a well-known Hamiltonian in
quantum optics. We consider a single three-level atom interacting with a cavity and we want to take into
account also the dissipative
terms, i.e., as in the previous Section, the spontaneous emission from the excited level $\gamma$ and the cavity
loss $\kappa$. Such
information can be embodied in the effective Hamiltonian (\ref{eq:Hoptcomp}) as follows:
\begin{equation}
H_{\rm opt}=\hbar[
-(\Delta+i\gamma)|e\rangle\langle{e}|-i\kappa{b^{\dagger}b} +\left(\Omega(t)|e\rangle\langle{g_1}|+gb|e\rangle\langle{g_3}|+h.c.\right)].
\end{equation}
We put the state vector $|\psi\rangle{\in}{\mathbb C}^3\otimes{\mathcal H}_{\rm cav}$ of the system in the form
\begin{equation}
|\psi\rangle=\sum_{m}(c_{e,m}|e,m\rangle+c_{g_1,m}|g_1,m\rangle+c_{g_3,m}|g_3,m+1\rangle)
\label{eq:vecstateopt}
\end{equation}
and by the Schr\"{o}dinger equation we found the equation of motion for the coefficients $c_{g_1,m}$, $c_{g_3,m}$ and
$c_{e,m}$:
\begin{eqnarray}
i\dot{c}_{e,m}&=&-(\Delta+i\gamma+i\kappa{m})c_{e,m}+{\Omega}c_{g_1,m}+g\sqrt{m+1}c_{g_3,m}, \nonumber\\
i\dot{c}_{g_1,m}&=&{\Omega}c_{e,m}-i\kappa{m}c_{g_1,m}, \nonumber\\
i\dot{c}_{g_3,m}&=&g\sqrt{m+1}c_{e,m}-i\kappa(m+1)c_{g_3,m}.
\label{eq:eqmotopt}
\end{eqnarray}
From the first equation one gets:
\begin{equation}
ic_{e,m}=\int_0^te^{i(\Delta+i\gamma+i\kappa{m})(t-\tau)}({\Omega}c_{g_1,m}+g\sqrt{m+1}c_{g_3,m})d\tau,
\end{equation}
then substituting the coefficient $c_{e,m}$ in the two last expressions of the Eq. (\ref{eq:eqmotopt}) with the
previous expression, imposing that $\Omega,g\sqrt{m+1}\ll\sqrt{\Delta^2+\gamma^2}$ and neglecting the terms
of higher order in $\Omega/\sqrt{\Delta^2+\gamma^2}$ and $g\sqrt{m+1}/\sqrt{\Delta^2+\gamma^2}$ one eventually
obtains the equations:
\begin{eqnarray}
i\dot{c}_{g_1,m}&=&\left(\frac{\Omega^2}{\Delta+i\gamma}-i\kappa{m}\right)c_{g_1,m}+
\frac{g\Omega\sqrt{m+1}}{\Delta+i\gamma}c_{g_3,m}, \nonumber\\
i\dot{c}_{g_3,m}&=&\left(\frac{g^2(m+1)}{\Delta+i\gamma}-i\kappa(m+1)\right)c_{g_3,m}+
\frac{g\Omega\sqrt{m+1}}{\Delta+i\gamma}c_{g_1,m}.
\label{eq:eqmotoptfin}
\end{eqnarray}
These equations can be equivalently derived starting from a
2-level system
 -- with internal states $|g_1\rangle$ and $|g_3\rangle$ -- interacting with the cavity mode by the effective,
 approximate Hamiltonian (\ref{eq:Heffopt}).

\subsection{Trapped atom in a cavity: Effective external Hamiltonian}\label{app:moteffH}

Let us write the non-Hermitian effective Hamiltonian, i.e., considering also the dissipative terms,
for a harmonically trapped 2-level atom inside a cavity QED:
\begin{equation}
H=(\omega_0-i\gamma)|e\rangle\langle{e}|+(\omega_c-i\kappa)b^{\dagger}b+\nu{a^{\dagger}a}\nonumber\\
+(\Omega|e\rangle\langle{g}|e^{i\omega_Lt}+h.c.)+\eta(a^{\dagger}+a)(gb|e\rangle\langle{g}|+h.c.).
\label{eq:H}
\end{equation}
Here, $\omega_0$ is the energy difference between the ground and the excited atomic level, $\omega_c$
and $b$ are respectively the energy and the annihilation operator for the cavity mode,  $\nu$
and $a$ are the energy and the annihilation operator for the harmonic motion, $\Omega$ and $\omega_L$ are
the Rabi frequency and the frequency of the laser light, $\eta$ is the Lamb-Dicke parameter, $g$ is
the dipole cavity-atom coupling constant and $\gamma$ and $\kappa$ are respectively the decay rate from the excited
state and the cavity loss. \\
From now on we do not consider in the equation the cavity loss, because its effect on the final
Hamiltonian is trivial -- indeed at the end it adds the term $-i\kappa b^{\dagger}b$: so we
re-introduce it just in the final Hamiltonian(s).
Making the canonical transformation $a\rightarrow{ae^{-i\nu{t}}}$, $b\rightarrow{be^{-i\omega_c{t}}}$ and
writing the state vector $|\psi\rangle\in{\mathbb C}^2\otimes{\mathcal H}_{\rm ext}\otimes{\mathcal H}_{\rm cav}$ of
the system in the following form:
\begin{equation}
|\psi\rangle=\sum_{n,m}(c_{e,n,m}|e,n,m-1\rangle+c_{g,n,m}|g,n,m\rangle),
\label{eq:vectorstate}
\end{equation}
we found the equation of motion for the coefficients $c_{e,n,m}$ and $c_{g,n,m}$ ($g,\Omega$ real):
\begin{eqnarray}
i\dot{c}_{e,n,m}&=&(\omega_0-i\gamma)c_{e,n,m}+{\Omega}e^{-i\omega_Lt}c_{g,n,m-1} 
+\eta{g}e^{-i\omega_ct}(\sqrt{nm}e^{i\nu{t}}c_{g,n-1,m}+\sqrt{(n+1)m}e^{-i\nu{t}}c_{g,n+1,m}), \nonumber\\
i\dot{c}_{g,n,m}&=&{\Omega}e^{i\omega_Lt}c_{e,n,m+1} 
+\eta{g}e^{i\omega_ct}(\sqrt{nm}e^{i\nu{t}}c_{e,n-1,m}+\sqrt{(n+1)m}e^{-i\nu{t}}c_{e,n+1,m}).
\end{eqnarray}
Integrating the first equation yields
\begin{equation}
ic_{e,n,m}=\int_0^te^{-i(\omega_0-i\gamma)(t-\tau)}[{\Omega}e^{-i\omega_L\tau}c_{g,n,m-1}
+\eta{g}e^{-i\omega_c\tau}(\sqrt{nm}e^{i\nu\tau}c_{g,n-1,m}+\sqrt{(n+1)m}e^{-i\nu\tau}c_{g,n+1,m})]d\tau
\end{equation}
then integrating by parts, considering the resonance condition $\omega_L+\nu=\omega_c$
and introducing the parameter $\Delta=\omega_L-\omega_0$ one gets the expression
\begin{eqnarray}
ic_{e,n,m}&=&e^{-i(\omega_0-i\gamma)t}\left[\frac{e^{-i(\Delta+i\gamma)\tau}}{-i(\Delta+i\gamma)}({\Omega}c_{g,n,m-1}
+\eta{g}(\sqrt{nm}c_{g,n-1,m}+\sqrt{(n+1)m}e^{-2i\nu\tau}c_{g,n+1,m}))\right]_0^t\nonumber\\
&-&\frac{i}{\Delta+i\gamma}\int_0^te^{-i(\Delta+i\gamma)\tau}\frac{d}{d\tau}({\Omega}c_{g,n,m-1}
+\eta{g}(\sqrt{nm}c_{g,n-1,m}+\sqrt{(n+1)m}e^{-2i\nu{\tau}}c_{g,n+1,m}))d\tau \nonumber
\end{eqnarray}
If in the previous expression we keep the terms up to the first order in
$\Omega/\sqrt{\Delta^2+\gamma^2}$, $\eta{g}/\sqrt{\Delta^2+\gamma^2}$, we obtain
\begin{equation}
c_{e,n,m}=\frac{e^{-i\omega_Lt}}{(\Delta+i\gamma)}({\Omega}c_{g,n,m-1}
+\eta{g}(\sqrt{nm}c_{g,n-1,m}+\sqrt{(n+1)m}e^{-2i\nu\tau}c_{g,n+1,m})).
\end{equation}
We can now substitute such an expression in the equation for $c_{g,n,m}$ and to first order
in $\eta$ we find
\begin{eqnarray}
i\dot{c}_{g,n,m}&=&\frac{\Omega^2}{\Delta+i\gamma}c_{g,n,m}\nonumber\\
&+&\frac{\eta{g}\Omega}{\Delta+i\gamma}(\sqrt{n(m+1)}c_{g,n-1,m+1}+\sqrt{(n+1)m}c_{g,n+1,m-1})\nonumber\\
&+&\frac{\eta{g}\Omega}{\Delta+i\gamma}(e^{-2i\nu{t}}\sqrt{(n+1)(m+1)}c_{g,n+1,m+1}+e^{2i\nu{t}}\sqrt{nm}c_{g,n-1,m-1}).\nonumber\\
\end{eqnarray}
The first term contains a Stark shift, namely $\Delta\Omega^2/(\Delta^2+\gamma^2)$, and a decoherence part
$\gamma\Omega^2/(\Delta^2+\gamma^2)$; the other terms, while leaving unchanged
the internal atomic states, involve the external (harmonic) atomic states and the cavity state. Thus it
is possible to write down an effective Hamiltonian for the external atomic degrees of freedom and the
cavity QED:
\begin{equation}
H_{\rm eff}=\frac{\eta{g}\Omega}{\Delta+i\gamma}(ab^{\dagger}+e^{2i\nu{t}}a^{\dagger}b^{\dagger}+h.c.)
-i\kappa{b^{\dagger}b}-i\frac{\gamma\Omega^2}{\Delta^2+\gamma^2}.
\end{equation}
If the RWA is applicable (i.e., in this case $\nu\gg{1/t}$ where $t$ is the time of the process we are
interested in) the previous expression takes the form Eq. (\ref{eq:HeffRWA}):
\begin{equation}
H_{\rm eff}^{\rm RWA}=\frac{\eta{g}\Omega}{\Delta+i\gamma}(ab^{\dagger}+a^{\dagger}b)-i\kappa{b^{\dagger}b}
-i\frac{\gamma\Omega^2}{\Delta^2+\gamma^2}.
\end{equation}

\section{Remarks on the adiabatic approximation}\label{app:pop}

The optical scheme after having introduced the pulse to compensate the light shift of the ground state $|g_1\rangle$ and
the motional scheme rely essentially on the same Hamiltonian. 
Thus we will study here the generic problem of adiabatic transfer in a system
driven by such a parametric Hamiltonian and we will eventually consider the actual expression of the parameters involved in
the specific scheme. The starting point is the 3-level coupling Hamiltonian
\begin{equation}
H=D|3\rangle\langle 3|+\hbar (G_1(t)|1\rangle\langle 3| + G_2(t)|2\rangle\langle 3| + h.c.), 
\label{eq:Hgen}
\end{equation}
where we have explicitely shown the time dependence of the coupling constants $G_i$, $i=1,2$. Note that for the motional
scheme $D=0$. 
The eigenfrequencies of the Hamiltonian Eq. (\ref{eq:Hgen}) are: $E_0=0$, $E_\pm=D\pm\sqrt{D^2+G_{\rm eff}^2}$,
where we defined $G_{\rm eff}=\sqrt{G_1^2+G_2^2}$.
The respective eigenstates can be written in the follwing form:
\begin{eqnarray}
|\psi_0(t)\rangle&=& G_{\rm eff}^{-1}(G_2|1\rangle-G_1|2\rangle), \nonumber \\
|\psi_\pm (t)\rangle&=&(E_\pm^2+G_{\rm eff}^2)^{-1/2}(G_1|1\rangle+G_2|2\rangle+E_\pm|3\rangle).
\end{eqnarray}
Troughout the paper we have called the zero-energy eigenstate $|\psi_0\rangle$ the dark state.
We prepare the system in the state $|1\rangle$. If at the beginning of the process $G_1/G_2=0$, 
we have that the dark state concides with such state. Then we suppose that the coupling constant $G_i$ 
are modified slowly -- adiabatically -- towards the ratio $G_1/G_2\rightarrow\infty$. 
The adiabatic theorem \cite{Galindo} tells us that the state of the system preceeds around the 
istantaneous eigenstate $|\psi_0(t)\rangle$, the asimptotic state being the state $|2\rangle$. The
population of the other (two) states, $P_\pm$, can be in principle evaluated in the adiabatic
approximation byusing the expression
\begin{equation}
P_\pm(t)=\left|\int_0^1e^{-it\alpha_{\pm,0}(\tau)}
\left\langle \psi_\pm(\tau)\left|\frac{d}{d\tau}\psi_0(\tau) \right. \right\rangle d\tau\right|^2,
\end{equation}
where $\alpha_{\pm,0}(t)=\int_0^t (E_\pm(t')-E_0(t')) dt'$.
An upper limit of the population is \cite{Messiah}
\begin{equation}
P_\pm \leq\left|\frac{\langle\psi_\pm|dH/dt|\psi_0\rangle}{\hbar(E_\pm-E_0)^2}\right|^2.
\end{equation}
Using the various expressions described before, one obtains
\begin{equation}
\frac{\langle\psi_\pm|dH/dt|\psi_0\rangle}{\hbar^2(E_\pm-E_0)^2}=
\left(1+\frac{D^2}{G_{\rm eff}^2}\right)^{-1}\left(1\pm \frac{D}{\sqrt{D^2+G_{\rm
eff}^2}}\right)^{-3/2}
\frac{\dot{G}_1G_2-G_1\dot{G}_2}{G_{\rm eff}^3}.
\label{eq:cpm}
\end{equation}
Considering $D=0$ and Gaussian-shaped coupling constant $G_1=G\exp[-(t/T-a)^2/\tau^2]$ and 
$G_2=G\exp[-(t/T+a)^2/\tau]^2$ in the previous expression we obtained for both
states
\begin{equation} 
P_\pm \leq P = \max_t\left\{\frac{a^2}{2 T^2 G^2 \tau^4}{\rm sech}^3
\left(\frac{2at}{T \tau^2}\right) e^\frac{t^2+a^2}{\tau^2}\right\}.
\end{equation}
The r.h.s. of the previous inequality diverges at $t\rightarrow\pm\infty$. We consider only a
finite interval -- the same we used in the numerical simulation of the transfer process -- where
anyway the condition on the initial and final value of the ratio $G_1/G_2$ are pretty well
satisfied. In such an interval the r.h.s. takes its maximum value when $G_1=G_2$, i.e., at the center of 
the interval (see also the numerical simulation). One obtains
\begin{equation} 
P_\pm \leq \frac{a^2}{2 T^2 G^2 \tau^4}e^{a^2/\tau^2}.
\label{eq:Ppm}
\end{equation}
When $D \neq 0$ the population for the two unwanted states is different. \\
In the text we are interested in the population of the leaky states to give a condition on the
rates $\gamma$ and $\kappa$. In the generic system considered here, this amounts
(as we will see soon explicitly) to evaluate the population $P_3$ of the state $|3\rangle$. 
From the expression for the eigenstate of the Hamiltonian we see that
\begin{equation}
P_3=\left|\frac{E_+}{\sqrt{G_{\rm eff}^2+E_+^2}}\langle \psi_+|\psi\rangle + 
\frac{E_-}{\sqrt{G_{\rm eff}^2+E_-^2}}\langle \psi_-|\psi\rangle \right|^2,
\end{equation}
where $|\psi\rangle$ is the state of the system during the adiabatic evolution.
For the optical scheme we have the identification 
$|1\rangle=|g_1\rangle_1|g_3\rangle_2|0\rangle_{\rm cav}$,
$|2\rangle=|g_3\rangle_1|g_1\rangle_2|0\rangle_{\rm cav}$ and
$|3\rangle=|g_3\rangle_1|g_3\rangle_2|1\rangle_{\rm cav}$ for the states and $G=g\Omega/\Delta$ for
the coupling constant peak value. In this case $D=g^2/\Delta \neq 0$; we have estimated
that, in our simulation, the term containing $D$ in Eq. (\ref{eq:cpm}) is of the order of 
the unity for $P_-$ and much smaller than 1 for
$P_+$. Thus an estimate of $P_3$ can be given as in the case $D=0$. In such a way, and
substituing the expression for $G$ in Eq. (\ref{eq:Ppm}), we
obtained Eq. (\ref{eq:pop1phopt}).
For the motional scheme we have the identification 
$|1\rangle=|1\rangle_1|0\rangle_2|0\rangle_{\rm cav}$,
$|2\rangle=|0\rangle_1|1\rangle_2|0\rangle_{\rm cav}$ and
$|3\rangle=|0\rangle_1|0\rangle_2|1\rangle_{\rm cav}$ for the states and $G=g\eta\Omega/\Delta$ for
the coupling constant peak value. Note that in this case $D=0$. Substituing $G$
in Eq. (\ref{eq:Ppm}) we obtained Eq. (\ref{eq:pop1phmot}).


\begin{thebibliography}{99}
\bibitem{scalability}
As we have shown, in principle, by choosing a system with
a sufficient number of levels, we can perform any $n$ qubit gate,
but such a scheme is not scalable up to large $n$ -- an
exponential number of ground states (namely $2^n+1$) would be required.
\bibitem{notice}
More constructively, in order to prove universality,  for a given
holonomic model one can  show how to find  suitable loops whose
holomomies are set of quantum gates that are known to be
universal.
\bibitem{norescond}
If it were not the case the term $\Delta_r|g_3\rangle\langle{g_3}|$, would be add to the Hamiltonian
(\ref{eq:Hoptcomp}).
\end{thebibliography}
\end{document}